\documentclass[3p,times,twocolumn]{elsarticle}
\biboptions{comma,sort&compress}
\usepackage{graphicx}
\usepackage[utf8]{inputenc}
\usepackage{here}
\usepackage{color}
\usepackage{ecrc}

\usepackage{tikz}
\usetikzlibrary{positioning,arrows}
\usetikzlibrary{decorations.pathmorphing}
\usetikzlibrary{decorations.markings}
\usepackage{pgfplots}
\usepackage{xparse}
\usetikzlibrary{positioning,arrows,patterns}
\usetikzlibrary{decorations.markings}
\usetikzlibrary{calc}

\volume{00}

\firstpage{1}

\journalname{Nuclear and Particle Physics Proceedings}

\runauth{}

\jid{nppp}

\jnltitlelogo{Nuclear and Particle Physics Proceedings}

\usepackage{amssymb}

\usepackage[figuresright]{rotating}

\def\bc{\begin{center}}
\def\ec{\end{center}}
\def\be{\begin{equation}}
\def\ee{\end{equation}}
\def\beqn{\begin{eqnarray}}
\def\eeqn{\end{eqnarray}}
\def\no{\nonumber}
\def\nn{\no\\}
\def\eqn#1{(\ref{#1})}
\def\ba{\begin{array}{c}}
\def\ea{\end{array}}
\def\bat{\begin{array}{cc}}
\def\bi{\begin{itemize}}
\def\ei{\end{itemize}}
\def\cA{{\cal A}}
\def\cL{{\cal L}}

\def\cO{{\cal O}}

\newcommand{\lsim}{~{}_{\textstyle\sim}^{\textstyle <}~}

\begin{document}

\begin{frontmatter}

\title{Updated Standard Model Prediction for $\varepsilon'/\varepsilon$ $^*$}

\cortext[cor0]{Talk given at 21th International Conference in Quantum Chromodynamics (QCD 18),  2 - 6 July 2018, Montpellier - FR}

\author[label1]{Hector Gisbert\corref{cor1}}
\ead{hector.gisbert@ific.uv.es}
\address[label1]{Departament de F\'\i sica Te\`orica, IFIC, Universitat de Val\`encia -- CSIC\\
Apt. Correus 22085, E-46071 Val\`encia, Spain}
\cortext[cor1]{Speaker.}
\author[label1]{Antonio Pich}
\ead{pich@ific.uv.es}

\pagestyle{myheadings}
\markright{ }
\begin{abstract}
A recent lattice evaluation of $\varepsilon'/\varepsilon$, finding a 2.1 $\sigma$ deviation from the experimental value, has revived the old debate about a possible $\varepsilon'/\varepsilon$ anomaly. The unfounded claims of a too low Standard Model prediction are based on incorrect estimates that neglect the long-distance re-scattering of the final pions in $K\to 2\pi$. In view of the current situation, we have recently updated the Standard Model calculation, including all known short- and long-distance contributions. Our result, $\mbox{Re}\left(\varepsilon'/\varepsilon\right) = (15 \pm 7)\cdot 10^{-4}$ \cite{Gisbert:2017vvj}, is in complete agreement with the experimental measurement.
\end{abstract}


\end{frontmatter}
\section{Introduction}

The CP violating ratio $\varepsilon'/\varepsilon$ constitutes a fundamental test for our understanding of flavour-changing phenomena. The present experimental world average \cite{Batley:2002gn,Lai:2001ki,Fanti:1999nm,Barr:1993rx,Burkhardt:1988yh,Abouzaid:2010ny,AlaviHarati:2002ye,AlaviHarati:1999xp,Gibbons:1993zq},
\be\label{eq:exp}
\mathrm{Re} \left(\varepsilon'/\varepsilon\right)\; =\;
(16.6 \pm 2.3) \cdot  10^{-4}\, ,
\ee
demonstrates the existence of direct CP violation in the decay transitions $K^0\to 2\pi$.

On the other hand, the theoretical prediction of $\varepsilon'/\varepsilon$ has been the subject of many debates. The first next-to-leading order (NLO) calculations \cite{Buras:1993dy,Buras:1996dq,Bosch:1999wr,Buras:2000qz,Ciuchini:1995cd,Ciuchini:1992tj} obtained Standard Model (SM) values one order of magnitude smaller than \eqn{eq:exp}. However, it was soon realized that the former SM predictions had missed an important  ingredient: the final-state interactions (FSI) of the two emitted pions \cite{Pallante:1999qf,Pallante:2000hk}. Once all relevant contributions were taken into account, the theoretical prediction was found to be in good agreement with the experimental value although with a large uncertainty of non-perturbative origin \cite{Pallante:2001he}.

Lattice QCD provides a suitable tool to face non-perturbative problems. However, the lattice efforts to explain the enhancement of the $\Delta I =1/2$ $K\to 2 \pi$ amplitude remained unsuccessful for many years, while attempts to estimate $\varepsilon'/\varepsilon$ were unreliable, sometimes even obtaining negative central values. The status of lattice simulations has improved considerably with the development of more sophisticated techniques and the increasing computer power. The RBC-UKQCD collaboration has achieved a successful calculation of the $\Delta I =3/2$ \ $K^+\to \pi^+\pi^0$ amplitude \cite{Blum:2011ng,Blum:2012uk,Blum:2015ywa}, and has recently obtained the first statistically-significant signal of the $\Delta I =1/2$ enhancement \cite{Boyle:2012ys}, 
in good agreement with the qualitative understanding achieved
long time ago with analytical techniques \cite{Pich:1995qp,Pich:1990mw,Jamin:1994sv,Bertolini:1997ir,Antonelli:1995nv,Antonelli:1995gw,Hambye:1998sma,Bardeen:1986vz,Buras:2014maa,Bijnens:1998ee}. 

The RBC-UKQCD group has also published a first estimate of
the direct CP-violation ratio, $\mathrm{Re}(\varepsilon'/\varepsilon) = (1.4 \pm 6.9) \cdot 10^{-4}$ \cite{Bai:2015nea,Blum:2015ywa}, which exhibits a 2.1 $\sigma$ deviation from the experimental value in  Eq. \eqn{eq:exp}. This result has brought back the old SM approaches predicting low values of $\varepsilon'/\varepsilon$ \cite{Buras:2015xba,Buras:2016fys,Buras:2015yba} and has triggered many new studies of possible contributions from physics beyond the SM \cite{Buras:2014sba,Buras:2015yca,Blanke:2015wba,Buras:2015kwd,Buras:2016dxz,Buras:2015jaq,Kitahara:2016otd,Kitahara:2016nld,Endo:2016aws,Endo:2016tnu,Cirigliano:2016yhc,Alioli:2017ces,Bobeth:2016llm,Bobeth:2017xry,Crivellin:2017gks,Chobanova:2017rkj,Bobeth:2017ecx,Endo:2017ums,Chen:2018dfc,Aebischer:2018quc,Aebischer:2018rrz,Haba:2018rzf,Matsuzaki:2018jui,Chen:2018vog,Chen:2018ytc,Haba:2018byj}. 
However, this discrepancy cannot be taken as evidence for new physics because the same lattice simulation fails to correctly reproduce the $(\pi\pi)_I$ phase shifts, which are a vital ingredient of the  $K^0\to\pi\pi$ calculation
and provide a quantitative test of the lattice results. While the extracted $I=2$ phase shift is only $1\,\sigma$ away from its physical value, the lattice analysis of Ref.~\cite{Bai:2015nea} finds a result for
$\delta_{0}$ which disagrees with the experimental value by $2.9\,\sigma$, a much larger discrepancy than the one quoted for $\varepsilon'/\varepsilon$. 
Efforts towards a better lattice understanding of the pion dynamics are under way and improved results are expected soon \cite{Kelly:CKM2018}.

Since the publication of the SM $\varepsilon'/\varepsilon$ prediction in Ref.~\cite{Pallante:2001he}, there have been a lot of improvements in
\begin{itemize}
\item The isospin-breaking corrections \cite{Cirigliano:2003nn,Cirigliano:2003gt,Cirigliano:2009rr}, which play a very important role in $\varepsilon'/\varepsilon$.
\item The quark masses, entering the large-$N_C$
determination of the penguin matrix elements, are nowadays known with much better precision \cite{Aoki:2016frl}.
\item The Low-Energy Constants (LECs), that appear in the Chiral Perturbation Theory ($\chi$PT) amplitudes, are better understood \cite{Ecker:1988te,Ecker:1989yg,Pich:2002xy,Cirigliano:2006hb,Kaiser:2007zz,Cirigliano:2004ue,Cirigliano:2005xn,RuizFemenia:2003hm,Jamin:2004re,Rosell:2004mn,Rosell:2006dt,Pich:2008jm,GonzalezAlonso:2008rf,Pich:2010sm,Bijnens:2014lea,Rodriguez-Sanchez:2016jvw,Ananthanarayan:2017qmx}.
\end{itemize}
The current situation makes mandatory to revise and update the analytical SM calculation of $\varepsilon'/\varepsilon$ \cite{Pallante:2001he}, which is already 17 years old. In the following, we summarize the new determination of $\varepsilon'/\varepsilon$ from Ref.~\cite{Gisbert:2017vvj}.

\section{Anatomy of $\boldmath \varepsilon'/\varepsilon$ }
\label{sec:anatomy}

Adopting the usual isospin decomposition, the kaon decay amplitudes can be expressed as
\begin{eqnarray}  
A(K^0 \to \pi^+ \pi^-) &=&  
\cA_{1/2} + {1 \over \sqrt{2}} \left( \cA_{3/2} + \cA_{5/2} \right) \nonumber\\
\; &=&\; 
 A_{0}\,  e^{i \chi_0}  + { 1 \over \sqrt{2}}\,    A_{2}\,  e^{i\chi_2 } \, ,
\no\\[2pt]
A(K^0 \to \pi^0 \pi^0) &=& 
\cA_{1/2} - \sqrt{2} \left( \cA_{3/2} + \cA_{5/2}  \right) 
\nonumber\\[2pt]
\; &=&\;
A_{0}\,  e^{i \chi_0}  - \sqrt{2}\,    A_{2}\,  e^{i\chi_2 }\, ,
\no\\[2pt]
A(K^+ \to \pi^+ \pi^0) &=&
{3 \over 2}  \left( \cA_{3/2} - {2 \over 3} \cA_{5/2} \right) \nonumber\\
\; &=&\;
{3 \over 2}\, A_{2}^{+}\,   e^{i\chi_2^{+}},
\label{eq:2pipar}
\end{eqnarray}
where the complex quantities $\cA_{\Delta I}$ are generated by the $\Delta I = \frac{1}{2},
\frac{3}{2},\frac{5}{2}$ components of the electroweak effective Hamiltonian, in the limit of isospin conservation. The $A_0$, $A_2$ and $A_2^+$ amplitudes are real and positive in the CP-conserving limit. Furthermore, in the isospin limit, $A_0$
and $A_2=A_2^+$ denote the decay amplitudes into $(\pi\pi)_I$ states with $I=0$ and 2, while the phase differences $\chi_0$ and $\chi_2=\chi_2^+$ are the corresponding S-wave scattering phase shifts. From the measured $K\to\pi\pi$ branching ratios, one gets \cite{Antonelli:2010yf}:
\begin{eqnarray}
A_0 &=& (2.704 \pm 0.001) \cdot 10^{-7} \mbox{ GeV}, \nn
A_2 &=& (1.210 \pm 0.002) \cdot 10^{-8} \mbox{ GeV}, \nn
\chi_0 - \chi_2 &=& (47.5 \pm 0.9)^{\circ}.
\label{eq:isoamps}
\end{eqnarray}
These numbers exhibit two important dynamical features that characterize the $K\to\pi\pi$ decay amplitudes:
\begin{enumerate}
\item Strong enhancement of the isoscalar amplitude with respect to the isotensor one, the so called ``$\Delta I =\frac{1}{2}$ rule'', 
\be
\hspace{1.cm}\omega \,\equiv\, 
\mathrm{Re} A_2 / \mathrm{Re} A_0\,\approx\, 1/22.
\ee
\item The S-wave re-scattering generates a large phase-shift difference between the $I= 0$ and $I=2$ partial waves.
This implies that 50\% of the $\cA_{1/2}/\cA_{3/2}$ ratio originates from the absorptive contribution:
\be\label{eq:AbsRat}
\hspace{1.1cm}\frac{\mathrm{Abs}\, (\cA_{1/2}/\cA_{3/2})}{\mathrm{Dis}\, (\cA_{1/2}/\cA_{3/2})}\, =\, 1.09\, .
\ee
\end{enumerate}
We would see later their strong implications on  $\varepsilon'/\varepsilon$.

In the presence of CP violation, the amplitudes $A_0$, $A_2$ and $A_2^+$ acquire imaginary parts. To first order in CP-violating quantities,
\begin{equation}
\varepsilon'\,  =\, 
- \frac{i}{\sqrt{2}} \: e^{i ( \chi_2 - \chi_0 )} \:\omega\;
\left[
\frac{\mathrm{Im} A_{0}}{ \mathrm{Re} A_{0}} \, - \,
\frac{\mathrm{Im} A_{2}}{ \mathrm{Re} A_{2}} \right] .
\label{eq:cp1}
\end{equation}
Thus, $\varepsilon'$ is suppressed by the ratio $\omega$ and $\varepsilon'/\varepsilon$ is approximately real since $\chi_2 - \chi_0 - \pi/2\approx 0$.
The CP-conserving amplitudes $\mathrm{Re} A_{I}$ are in general fixed to their experimental values, given in Eq.~(\ref{eq:isoamps}), in order to reduce the theoretical uncertainty. A theoretical calculation is then only needed for $\mathrm{Im} A_{I}$. 

Eq.~\eqn{eq:cp1} contains a subtle numerical balance between the two isospin contributions, making the result very sensitive to the values of the CP-violating amplitudes. Naive estimates of $\mathrm{Im} A_{I}$ result in a strong cancellation between the two terms, leading to unrealistically low values of $\varepsilon'/\varepsilon$ \cite{Buras:1993dy,Buras:1996dq,Bosch:1999wr,Buras:2000qz,Ciuchini:1995cd,Ciuchini:1992tj,Buras:2015xba,Buras:2016fys,Buras:2015yba}, as we show in section~\ref{sec:strongcancellation}.

Isospin breaking effects are very important in $\varepsilon'/\varepsilon$, due to the large ratio $1/\omega$ \cite{Cirigliano:2003nn,Cirigliano:2003gt,Cirigliano:2009rr}. Including isospin violation, $\mathrm{Re}(\varepsilon'/\varepsilon)$ can be written as
\be\label{eq:epsp_simp}
\mbox{}\hskip -.6cm
\mathrm{Re}\Bigl(\frac{\varepsilon'}{\varepsilon}\Bigr) \, = \, - \frac{\omega_+}{\sqrt{2}\, |\varepsilon|}  \, \left[
\frac{\mathrm{Im} A_{0}^{(0)} }{ \mathrm{Re} A_{0}^{(0)} }\,
\left( 1 - \Omega_{\rm eff} \right) - \frac{\mathrm{Im} A_{2}^{\rm emp}}{ \mathrm{Re}
  A_{2}^{(0)} } \right]  , \;\;
\ee
where $\mathrm{Im} A_2^{\rm emp}$ contains the $I=2$ contribution from the electromagnetic penguin operator, the superscript $(0)$ denotes the isospin limit, and $\omega_+ \equiv \mathrm{Re} A_{2}^{+}/\mathrm{Re} A_{0}$ is directly extracted from \eqn{eq:isoamps}. The parameter
\be
\Omega_{\rm eff} \; =\; \Omega_{\rm IV} - \Delta_0 - f_{5/2}
\; =\; (6.0\pm 7.7) \cdot 10^{-2}
\ee
incorporates the computed isospin-breaking corrections~\cite{Cirigliano:2003nn,Cirigliano:2003gt}.

\section{Theoretical framework}
The physical origin of $\varepsilon'/\varepsilon$ is at the electroweak scale where all the flavour-changing processes are described in terms of  quarks and gauge bosons. Due to the presence of very different mass scales ($M_\pi < M_K \ll M_W$), the gluonic corrections  to the $K\to\pi\pi$ amplitudes are amplified with large logarithms that can be summed up using the Operator Product Expansion (OPE) and the renormalization group equations (RGEs), all the way down to scales $\mu<m_c$. Finally, one gets an effective Lagrangian  defined in the three-flavour theory \cite{Buchalla:1995vs},
\be\label{eq:Leff}
\cL_{\mathrm{eff}}^{\Delta S=1}\, =\, - \frac{G_F}{\sqrt{2}}\,
 V_{ud}^{\phantom{*}}V^*_{us}\,  \sum_{i=1}^{10}
 C_i(\mu) \, Q_i (\mu)\, ,
\ee
which is a sum of local operators weighted by short-distance coefficients $C_i(\mu)$ that depend on the heavy  masses ($\mu>M$) and CKM parameters. The Wilson coefficients $C_i(\mu)$ are known at NLO \cite{Buras:1991jm,Buras:1992tc,Buras:1992zv,Ciuchini:1993vr}. This includes all corrections of $\cO(\alpha_s^n t^n)$ and $\cO(\alpha_s^{n+1} t^n)$, where
$t\equiv\log{(M_1/M_2)}$ refers to the logarithm of any ratio of
heavy mass scales $M_1,M_2\geq\mu$. Some next-to-next-to-leading-order (NNLO) corrections are already known \cite{Buras:1999st,Gorbahn:2004my} and efforts towards a complete short-distance calculation at the NNLO are currently under way \cite{Cerda-Sevilla:2016yzo}.

\begin{figure}[t]\centering
\setlength{\unitlength}{0.65mm} 
\resizebox{.96\totalheight}{!}{
\begin{picture}(156,122)
\put(0,0){\makebox(156,120){}}
\thicklines
\put(8,111){\makebox(25,10){Energy}}
\put(43,111){\makebox(42,10){Fields}}
\put(101,111){\makebox(52,10){Effective Theory}}
\put(5,110){\line(1,0){146}} {
\put(8,76){\makebox(25,30){$M_W$}}
\put(47,78){\framebox(34,25){
   $\ba W, Z, \gamma, G_a \\  \tau, \mu, e, \nu_i \\ t, b, c, s, d, u \ea $}}
\put(101,76){\makebox(52,30){Standard Model}}

\put(8,38){\makebox(25,20){$\lsim m_c$}}
\put(47,40){\framebox(34,16){
 $\ba  \gamma, G_a  \, ;\, \mu ,  e, \nu_i \\ s, d, u \ea $}}
\put(101,38){\makebox(52,20){$\cL_{\mathrm{QCD}}^{N_f=3}$,
             $\cL_{\mathrm{eff}}^{\Delta S=1,2}$}}

\put(8,0){\makebox(25,20){$m_K$}}
\put(47,2){\framebox(34,16){
 $\ba\gamma \; ;\; \mu , e, \nu_i  \\  \pi, K,\eta  \ea $}}
\put(101,0){\makebox(52,20){$\chi$PT}}
\linethickness{0.3mm}
\put(64,36){\vector(0,-1){15}}
\put(64,74){\vector(0,-1){15}}
\put(69,65){OPE}
\put(69,27){$N_C\to\infty $}}    
\end{picture}
}
\vskip -.5cm\mbox{}
\caption{Evolution from $M_W$ to the kaon mass scale.
  \label{fig:eff_th}}
\end{figure}
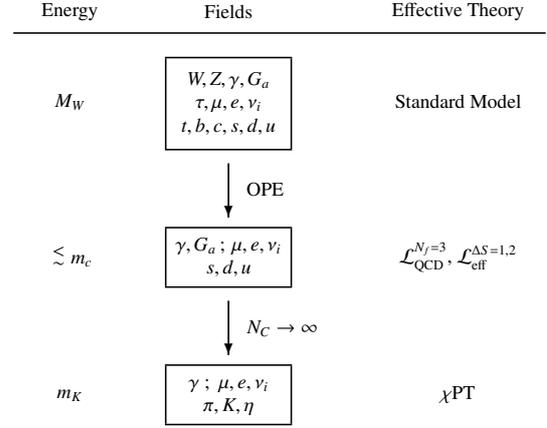

Below the resonance region, where perturbation theory no longer works, we can use symmetry considerations to define another effective field theory in terms of the QCD Goldstone bosons ($\pi$, $K$, $\eta$). $\chi$PT  describes the pseudoscalar octet dynamics through a perturbative expansion in powers of momenta and quark masses over the chiral symmetry breaking scale $\Lambda_\chi\sim 1$~GeV \cite{Weinberg:1978kz,Gasser:1983yg,Gasser:1984gg}. At lowest order, the most general effective bosonic Lagrangian with the same SU$(3)_L\otimes \:$SU$(3)_R$ transformation properties as $\cL_{\mathrm{eff}}^{\Delta S=1}$ contains three terms \cite{Cirigliano:2011ny}:
\begin{eqnarray}\label{eq:LchiPT}
\cL_{2}^{\Delta S=1}\:=\:G_8\:\cL_{8}\:+\:G_{27}\:\cL_{27}\:+\:G_8\:g_{\mathrm{ewk}}\:\cL_{\mathrm{ewk}}\, .
\end{eqnarray}
Thus, $\cL_{2}^{\Delta S=1}$ determines the $K^0\to\pi\pi$ amplitudes at $\mathcal{O}(p^2)$ in terms of the LECs $G_8$, $G_{27}$ and $G_8\, g_{\mathrm{ewk}}$. 
A first-principle computation of these three LECs requires to perform a matching between the short-distance and effective  Lagrangians in Eqs.~(\ref{eq:Leff}) and (\ref{eq:LchiPT}). This can be easily done in the limit of an infinite number of quark colours, where the four-quark operators factorize into currents with well-known chiral realizations. Since the large-$N_C$ limit is only applied in the matching between the two effective field theories, the only missing contributions are $1/N_C$ corrections that are not enhanced by any large logarithms. Figure~\ref{fig:eff_th} shows schematically the chain of effective theories entering the analysis of the kaon decay dynamics.

\section{Strong cancellation in simplified analysis}\label{sec:strongcancellation}

The CP-odd amplitudes in Eq.~(\ref{eq:epsp_simp}) are dominated by the penguin operators $Q_6$ and $Q_8$, due to their chiral enhancement. Taking only into account these two operators and ignoring all others in the estimation of $\mathrm{Im} A_{I}$, one finds \cite{Buras:1987wc,Buras:1985yx}: 
\beqn \label{eq:A0Q6}
\left.\mathrm{Im} A_{0}\right|_{Q_6} & =& 
 \mathfrak{X}_6 \:4\sqrt{2}\: \left(F_K - F_\pi\right)\:    B_6^{(1/2)} ,
\\[5pt]\label{eq:A2Q8}
\left.\mathrm{Im} A_{2}\right|_{Q_8} & =&
 -   \:\mathfrak{X}_8\:2\:
F_\pi\:    B_8^{(3/2)} ,
\eeqn
where $\mathfrak{X}_i\equiv \frac{G_F}{\sqrt{2}}\: A^2\:\lambda^5\:\eta\:
  \:y_i(\mu)\: \left[\frac{M_K^2}{m_d(\mu) + m_s(\mu)}\right]^2$, $F_K$  is the kaon decay constant and the factors $B_6^{(1/2)}$ and $B_8^{(3/2)}$ parametrize the deviations of the true hadronic matrix elements from their large-$N_C$ approximations. Notice, in the definition of $\mathfrak{X}_i$, that the short-distance scale of $y_6(\mu)$ and $y_8(\mu)$ is cancelled by the scale dependence of the quark masses since, at $N_C\to\infty$, the only elements of the anomalous dimension matrix $\gamma_{ij}$  that survive are $\gamma_{66}$ and $\gamma_{88}$ \cite{Bardeen:1986uz}. This nice scale cancellation illustrates how the product of two colour-singlet quark currents factorizes and, in addition, that the product $m_q \bar q q$ is renormalization-scale invariant.

Introducing this rough prediction of the CP-odd amplitudes in Eq.~\eqn{eq:epsp_simp}, one obtains
\be\label{eq:epsilonp_naive}
\mbox{}\hskip -.7cm
\mathrm{Re}\Bigl(\frac{\varepsilon'}{\varepsilon}\Bigr) \, \approx \, 
2.2\cdot 10^{-3}
\left\{ B_6^{(1/2)} \left( 1 - \Omega_{\rm eff} \right) - 0.48\, B_8^{(3/2)}
\right\}  .\;
\ee
With $B_6^{(1/2)} = B_8^{(3/2)} = 1$ (large-$N_C$ values) and $\Omega_{\rm eff}= 0.06$, the naive estimation gives
$\mathrm{Re}(\varepsilon'/\varepsilon) \approx 1.0\cdot 10^{-3}$ as the order of magnitude for the SM prediction.
An interesting observation is the delicate cancellation among the different terms in~\eqn{eq:epsilonp_naive} which makes the final number very sensitive to the chosen inputs \cite{Bertolini:1995tp,Bertolini:1997nf,Bertolini:1998vd,Bertolini:2000dy,Hambye:1999yy}.  
With the values adopted in Ref.~\cite{Buras:2015yba}, $B_6^{(1/2)} = 0.57$, $B_8^{(3/2)} = 0.76$ and $\Omega_{\rm eff}= 0.15$, one obtains $\mathrm{Re}(\varepsilon'/\varepsilon) \approx 2.6\cdot 10^{-4}$, which is an order of magnitude smaller and in clear conflict with the experimental value in Eq.~\eqn{eq:exp}.

However, these simplified estimates neglect completely the strong FSI present in $K^0\to \pi\pi$ decays, and miss the very large absorptive contribution giving rise to the measured phase-shift difference. The two relevant decay amplitudes get large logarithmic corrections from pion loops \cite{Pallante:1999qf,Pallante:2000hk,Pallante:2001he} that can be rigorously calculated using the usual $\chi$PT methods. They turn out to be positive for $\left. A_{0}\right|_{Q_6}$, while negative for $\left. A_{2}\right|_{Q_8}$. Consequently, the numerical cancellation in Eq.~\eqn{eq:epsilonp_naive} disappears and one gets  a sizeable enhancement of the SM prediction for $\varepsilon'/\varepsilon$, in good agreement with its experimental value.

\section{{\boldmath $K\to\pi\pi$} amplitudes in {\boldmath $\chi$}PT}

With the Lagrangian~(\ref{eq:LchiPT}), the kaon decay amplitudes  are easily obtained at $\mathcal{O}(p^2)$ through a simple perturbative calculation in $\chi$PT. One only needs to consider tree-level Feynman diagrams with one insertion of $\cL_2^{\Delta S=1}$. Assuming isospin conservation, the $\cA_{\Delta I}$ amplitudes defined in Eq.~\eqn{eq:2pipar} are given by \cite{Pallante:2001he,Cirigliano:2003gt} 
\begin{eqnarray}
\cA_{1/2} &\!\!\! = & \!\!\! - \sqrt{2}\, G_8 F\, \Big[  \left( M_{K}^2 - M_{\pi}^2 \right)
- {2 \over 3}\, F^2  e^2 g_{\rm ewk} \Big]\nonumber\\
&\!\!\! &\!\!\! - {\sqrt{2} \over 9}\, G_{27} F \left( M_{K}^2 - M_{\pi}^2 \right)  ,
\nn
\cA_{3/2} &\!\!\! = &\!\!\! \!\mbox{} -
{10 \over 9}  G_{27} F \left( M_{K}^2 - M_{\pi}^2 \right)
+  {2 \over 3} G_8 F^3 e^2   g_{\mathrm{ewk}}\, ,\nn
\cA_{5/2} &\!\!\! = &\!\!\! 0\, .
\label{eq:LOamp}
\end{eqnarray}
From the measured amplitudes in Eq.~\eqn{eq:isoamps}, one immediately obtains the tree-level determinations $g_8 = 5.0$ and $g_{27} = 0.25$ for the octet and 27-plet chiral couplings, respectively, with the normalization 
$G_{8,27}\:\equiv\:-\:\frac{G_F}{\sqrt{2}}\:V_{ud}\:V^*_{us}\:g_{8,27}$.
The large numerical difference between these two LECs just reflects the 
smallness of the measured ratio $\omega\approx 5 \sqrt{2}\, g_{27}/(9\, g_8)$.

\begin{figure}[t]\centering
\begin{center}
\begin{minipage}[t]{0.5\textwidth}
\centering
\tikzset{
  goldstone/.style={draw=black}
}

\NewDocumentCommand\semiloop{O{black}mmmO{}O{above}}
{
\draw[#1] let \p1 = ($(#3)-(#2)$) in (#3) arc (#4:({#4+180}):({0.5*veclen(\x1,\y1)})node[midway, #6] {#5};)
}

\mbox{}\hskip -.7cm
\begin{tikzpicture}[line width=1.5 pt,node distance=1cm and 1cm]
\coordinate[ ] (v1);
\coordinate[right=1.5cm of v1] (v4);
\coordinate[above right=of v4,label=right:$\pi$] (f1);
\coordinate[below right=of v4,label=right:$\pi$] (f2);
\coordinate[ left =of v1,label=left :$K^0$] (e2);
\draw[goldstone] (v1) -- (e2);
\draw[goldstone] (v4) -- (f1);
\draw[goldstone] (f2) -- (v4);
\semiloop[goldstone]{v1}{v4}{0}[$ $];
\semiloop[goldstone]{v4}{v1}{180}[$ $][below];
\coordinate[above right=1.0cm and -0.5cm of v4,label=left:$ $];
\coordinate[above right=-1.0cm and -0.5cm of v4,label=left:$ $];
\fill[red] (v1) circle (.1cm);
\end{tikzpicture}
\end{minipage}
\end{center}
\vskip -.5cm
\caption{One-loop $K\rightarrow\pi\pi$ topology with an absorptive cut.}
\label{fig:1-loop}
\end{figure}
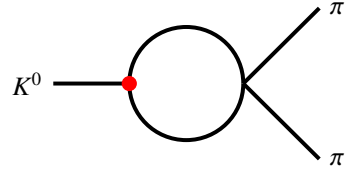

At LO in $\chi$PT, the phase shifts are predicted to be zero, because they are generated through loop diagrams with $\pi\pi$ absorptive cuts. Figure~\ref{fig:1-loop} displays the only one-loop topology contributing to the absorptive amplitudes.
The large value of the measured phase-shift difference in Eq.~\eqn{eq:isoamps} indicates a very large absorptive contribution. Analyticity relates the absorptive and dispersive parts of the one-loop diagram, which implies that the dispersive correction is also very large. 
A proper calculation of chiral loop corrections is then compulsory in order to obtain a reliable prediction for $\varepsilon'/\varepsilon$.

The incorrect estimates, claiming small SM values of $\varepsilon'/\varepsilon$ \cite{Buras:2015xba,Buras:2016fys,Buras:2015yba}, are flawed because they totally ignore the presence of absorptive cuts. They are based on (model-dependent) real  $K\to\pi\pi$ amplitudes that fail to comply with the experimental constraint in Eq.~\eqn{eq:AbsRat}.

At the NLO in $\chi$PT, the $\cA_{\Delta I}$ amplitudes can be written in the form
\begin{eqnarray}
\mbox{}\hskip -.2cm
\mathcal{A}_{\Delta I} \, &\!\!\! =&\!\!\! -\,
G_8 F_\pi\,\Bigl\{ (M_K^2-M_\pi^2)\,\mathcal{A}_{\Delta I}^{(8)} -e^2 F_{\pi}^2 \, g_{\mathrm{ewk}}\,\mathcal{A}_{\Delta I}^{(g)}\Bigr\}
\nonumber\\
&&\!\!\! -\, G_{27} F_\pi\, (M_K^2-M_\pi^2)\,\mathcal{A}_{\Delta I}^{(27)}\, ,
\label{amplitudegeneral}
\end{eqnarray}
where $\mathcal{A}_{\Delta I}^{(8)}$ and $\mathcal{A}_{\Delta I}^{(27)}$ represent the octet and 27-plet components, and $\mathcal{A}_{\Delta I}^{(g)}$ contains the electroweak penguin contributions. Moreover, these quantities can be further decomposed as
\be 
\cA_{\Delta I}^{(X)}\, =\, a_{\Delta I}^{(X)} \,\left[ 1 + \Delta_L\cA_{\Delta I}^{(X)} + \Delta_C\cA_{\Delta I}^{(X)}\right]\, ,
\ee
with $a_{\Delta I}^{(X)}$
the tree-level contributions, $\Delta_L\cA_{\Delta I}^{(X)}$ the one-loop chiral corrections and $\Delta_C\cA_{\Delta I}^{(X)}$ the NLO local corrections at $\mathcal{O}(p^4)$. 
The numerical values of  the different $\cA_{1/2}^{(X)}$ and $\cA_{3/2}^{(X)}$ components are displayed in tables~\ref{tab:Table12} and \ref{tab:Table32}, respectively. 

\begin{table}[hb] 
\renewcommand*{\arraystretch}{1.3}
\begin{center}
\begin{tabular}{|c|c|c|c|c|}\hline 
X & $\! a_{1/2}^{(X)}\! $ &
$\Delta_{L} \mathcal{A}_{1/2}^{(X)} $ &
$[\Delta_{C} \mathcal{A}_{1/2}^{(X)}]^+ $  & 
$[\Delta_{C} \mathcal{A}_{1/2}^{(X)}]^- $ \\
\hline   
\hline 
8  & $\sqrt{2}$   & $\! 0.27 +  0.47\,  i\! $   &
$\! \phantom{-}0.01 \pm 0.05\! $   &  $\! \phantom{-}0.02 \pm 0.05\! $  
\\
\hline 
g  & $ \! \frac{2 \sqrt{2}}{3}\! $  &  $\! 0.27 +  0.47\, i\! $ & 
$\! -0.19 \pm 0.01\! $  & $\! -0.19 \pm 0.01\! $ 
\\
\hline 
$\!\! 27\!\!$  &  $\frac{\sqrt{2}}{9}$   &  $\! 1.03 + 0.47\, i\! $    & 
$\! \phantom{-}0.01 \pm 0.63\! $ & $\! \phantom{-}0.01  \pm 0.63\! $ 
\\
\hline   
\end{tabular}
\caption{Numerical predictions for the $\mathcal{A}_{1/2}$ components. The local NLO correction to the CP-even ($[\Delta_{C} \mathcal{A}_{1/2}^{(X)}]^+$) and CP-odd ($[\Delta_{C} \mathcal{A}_{1/2}^{(X)}]^-$) amplitudes is only different in the octet case.} 
\label{tab:Table12} 
\vskip .5cm
%
\begin{tabular}{|c|c|c|c|}\hline 
X & $a_{3/2}^{(X)}$ &
$\Delta_{L} \mathcal{A}_{3/2}^{(X)} $ & $\Delta_{C} \mathcal{A}_{3/2}^{(X)} $
\\
\hline   
\hline 
g  & $ \frac{2}{ 3}$  &  $-0.50 -  0.21\; i$  & $-0.19 \pm 0.19$      
\\
\hline 
27  &  $\frac{10}{9}$  &  $-0.04 - 0.21\; i$    & 
$\phantom{-}0.01 \pm 0.05$    
\\
\hline   
\end{tabular}
\caption{Numerical predictions for the $\mathcal{A}_{3/2}$ components. }  
\label{tab:Table32} 
\end{center} 
\end{table}

The absorptive chiral corrections are large and positive for the $\Delta I=1/2$ amplitudes and much smaller and negative for $\Delta I=3/2$. Furthermore, they do not depend on the chiral renormalization scale $\nu_\chi$. Besides, table~\ref{tab:Table12} shows a huge dispersive one-loop correction to the $\cA^{(27)}_{1/2}$ amplitude. However, since Im$(g_{27})=0$, the 27-plet components do not contribute to the CP-odd amplitudes and, therefore, do not introduce any uncertainty in the final numerical value of Im$A_0$.

The relevant NLO loop corrections for $\varepsilon'/\varepsilon$ are $\Delta_L\cA_{1/2}^{(8)}$ and $\Delta_L\cA_{3/2}^{(g)}$. The first one generates a significant enhancement of Im$A_0$, $|1 +\Delta_L\cA_{1/2}^{(8)}|\approx 1.35$, while the second one produces a suppression in Im$A_2^{\mathrm{emp}}$, $|1 +\Delta_L\cA_{3/2}^{(g)}|\approx 0.54$. Consequently, the numerical cancellation between the $I=0$ and $I=2$ terms in Eq.~\eqn{eq:epsilonp_naive} is completely destroyed by the chiral loop corrections.

Tables~\ref{tab:Table12} and \ref{tab:Table32} show also the numerical predictions for the NLO local corrections $\Delta_C\cA_{\Delta I}^{(X)}$, which have been estimated in the large-$N_C$ limit. The dependence with $\nu_\chi$ (absent at large $N_C$) is our main source of uncertainty. In order to estimate the errors, we have varied $\nu_\chi$ between 0.6 and 1 GeV in the corresponding loop contributions $\Delta_L\cA_{\Delta I}^{(X)}$. In addition, we have taken the uncertainty associated to the short-distance scale varying $\mu$ between $M_\rho$ and $m_c$, but the impact on the $\Delta_C\cA_{\Delta I}^{(X)}$ corrections is negligible compared with the $\nu_\chi$ uncertainty. The most significant local corrections for $\varepsilon'/\varepsilon$ are
$[\Delta_C\cA_{1/2}^{(8)}]^-$ and $\Delta_C\cA_{3/2}^{(g)}$; nevertheless, they are much smaller than the loop contributions. 

\section{\boldmath The SM prediction for $\varepsilon'/\varepsilon$}

Taking into account all computed corrections in Eq.~\eqn{eq:epsp_simp}, our SM prediction for $\varepsilon'/\varepsilon$ is
\beqn\label{eq:finalRes}
\mbox{Re}\left(\varepsilon'/\varepsilon\right)&\!\!\! =&\!\!\!
\left(15\pm 2_{\mu}\pm 2_{m_s} \pm 2_{\Omega_\mathrm{eff}}\pm 6_{1/N_C}\right) \times 10^{-4}
\no\\[5pt] &\!\!\! =&\!\!\! 
\left(15\pm 7\right) \times 10^{-4}\, .
\eeqn
The first uncertainty has been estimated by varying the short-distance renormalization scale $\mu$ between $M_\rho$ and $m_c$. The second error shows the sensitivity to the strange quark mass, within its allowed range, while the third one displays the uncertainty from the isospin-breaking parameter $\Omega_{\mathrm{eff}}$. The last error is our dominant source of uncertainty and reflects our ignorance
about $1/N_C$-suppressed contributions that we have missed in the matching process.

\begin{figure}[t]\centering
\includegraphics[scale=.5]{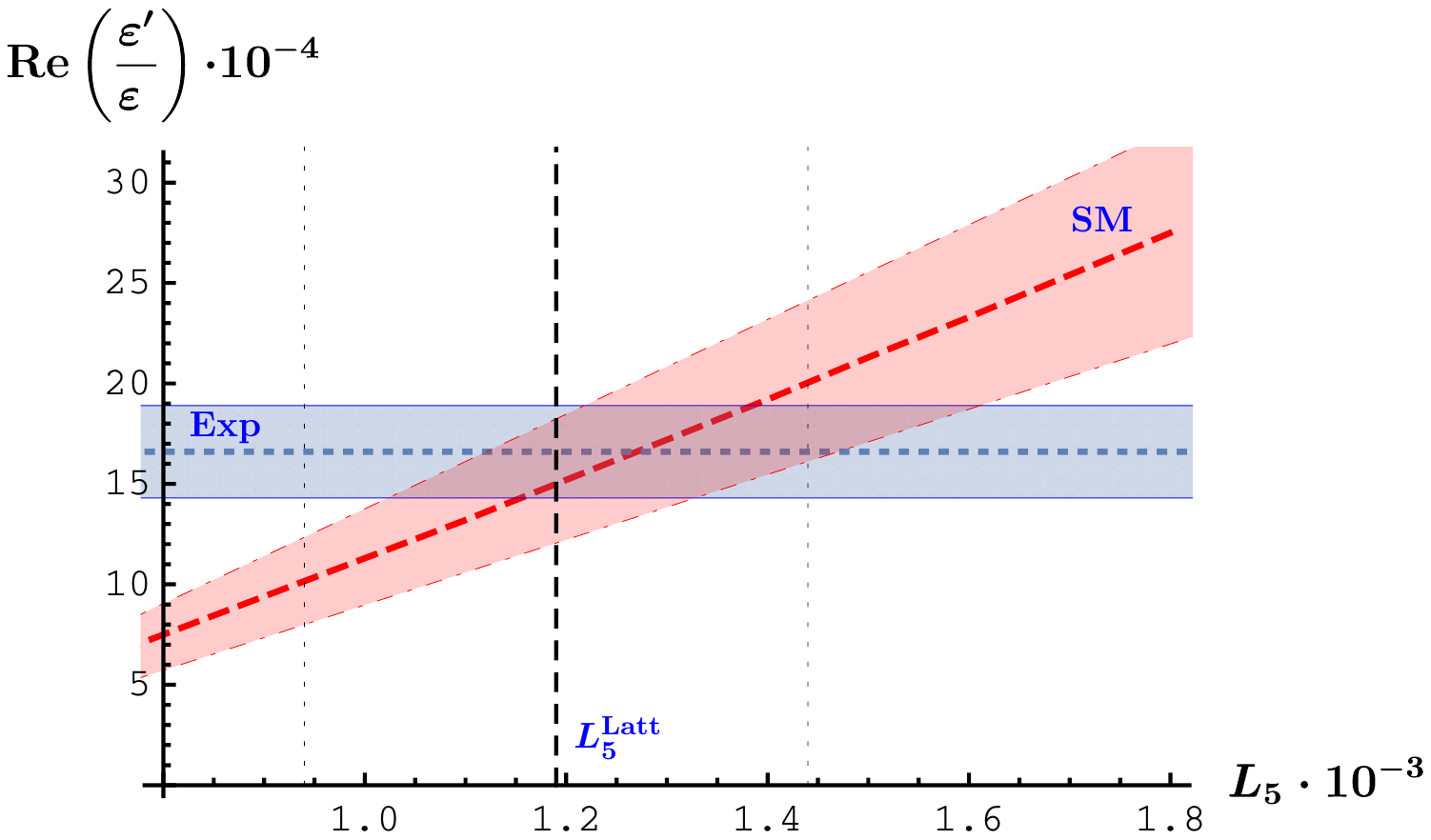}
\caption{SM prediction for $\varepsilon'/\varepsilon$ as function of $L_5$
(red dashed line) with $1\,\sigma$ errors (oblique band).
The horizontal blue band displays the experimentally measured value with $1\,\sigma$ error bars. The dashed vertical line shows the current lattice determination of $L_5^r(M_\rho)$.}
\label{fig:epsp}
\end{figure}

In figure~\ref{fig:epsp}, we plot the prediction for $\varepsilon'/\varepsilon$ as function of the $\chi$PT coupling $L_5$, which clearly shows a strong dependence on this parameter. The experimental $1\,\sigma$ range is indicated by the horizontal band, while the dashed vertical lines display the current lattice 
determination of $L_5^r(M_\rho)$. The measured value of $\varepsilon'/\varepsilon$ is nicely reproduced with the preferred lattice inputs.

\section{Final remarks}

Our SM prediction for $\varepsilon'/\varepsilon$ is in perfect agreement with the measured experimental value. We have shown the important role of FSI in $K^0\to\pi\pi$. When $\pi\pi$ re-scattering corrections are taken into account, the numerical cancellation between the $Q_6$ and $Q_8$ terms in Eq.~\eqn{eq:epsilonp_naive} is completely destroyed because of the positive enhancement of the $Q_6$ amplitude and the negative suppression of the $Q_8$ contribution. Once these important corrections are included, the contributions from other four-quark operators to $\mathrm{Im} A_0^{(0)}$ and $\mathrm{Im} A_2^{\mathrm{emp}}$ become numerically less relevant, since  the cancellation is no longer operative.

The claims \cite{Buras:2015xba,Buras:2016fys,Buras:2015yba} of a flavour anomaly in $\varepsilon'/\varepsilon$  originate in naive approximations that overlook the important role of pion chiral loops. These incorrect estimates are using simplified ansatzs for the $K\to\pi\pi$ amplitudes, without any absorptive contributions, in complete disagreement with the strong experimental evidence of a very large phase shift difference.

The recent lattice results look quite encouraging, since it is the first time that a clear signal of the $\Delta I=1/2$ enhancement seems to emerge from lattice data \cite{Boyle:2012ys}. RBC-UKQCD has obtained a first numerical estimate of $\varepsilon'/\varepsilon$ with a quite small central value, 2.1 $\sigma$ lower than the experimental measurement. However, the same lattice simulation finds a $(\pi\pi)_{I=0}$ phase shift $2.9\,\sigma$ away from its physical value, which indicates  that these results are still in a very premature stage. Substantial improvements, with much larger statistics and a better control of the final pion dynamics, are expected soon.

Our prediction of $\varepsilon'/\varepsilon$ agrees well with the measured value, providing a qualitative confirmation of the SM mechanism of CP violation. Although the theoretical error is still large, improvements can be achieved  in the next years via a combination of analytical calculations, numerical simulations and data analyses. For instance,
\begin{itemize}
\item A computation of the Wilson coefficients at NNLO is currently been performed \cite{Cerda-Sevilla:2016yzo}. 
\item The isospin-breaking effects are vital for a correct $\varepsilon'/\varepsilon$ prediction. A complete re-analysis with updated inputs is currently under way \cite{VincenzoEtAl}.
\item The $\cO(e^2p^0)$ coupling $g_8\, g_{ewk}$ can be expressed as a dispersive integral over the hadronic vector and axial-vector spectral functions. 
The $\tau$ decay data can then be used to perform a direct determination of this LEC. A new phenomenological analysis is close to being finalized \cite{Antonio}.

\item The dominant $\chi$PT corrections originate from large chiral logarithms. A reliable estimate of higher-order contributions should be feasible either through explicit two-loop calculations or with dispersive techniques \cite{Pallante:1999qf,Pallante:2000hk,Pallante:2001he,Buchler:2001nm,Buchler:2005xn}.

\item A matching calculation of the weak LECs at NLO in $1/N_C$ remains a very challenging task. A fresh view to previous attempts \cite{Bertolini:1995tp,Bertolini:1997nf,Bertolini:1998vd,Bijnens:2000im,Hambye:2003cy,Pich:1995qp,Pich:1990mw,Jamin:1994sv,Bertolini:1997ir,Antonelli:1995nv,Antonelli:1995gw,Hambye:1998sma,Bardeen:1986vz,Buras:2014maa,Bijnens:1998ee}, with a modern perspective, could suggest new ways to face this unsolved problem.

\item Lattice QCD simulations are expected to provide new improved data on $K^0\to\pi\pi$ transitions in the next years \cite{Kelly:CKM2018,Feng:2017voh}. Combined with appropriate $\chi$PT techniques, a better control of systematic uncertainties could be achieved. 
\end{itemize}

\section*{Acknowledgements}

We want to thank the organizers for their effort to make this conference such a successful event. We also thank Vincenzo Cirigliano, Gerhard Ecker, Elvira Gámiz, Helmut Neufeld, Jorge Portolés and Antonio Rodríguez Sánchez  for useful discussions. This work has been supported in part by the Spanish State Research Agency and ERDF funds from the EU Commission [Grants FPA2017-84445-P and FPA2014-53631-C2-1-P], by Generalitat Valenciana [Grant Prometeo/2017/053] and by the Spanish Centro de Excelencia Severo Ochoa Programme [Grant SEV-2014-0398]. The work of H.G. is supported by a FPI doctoral contract [BES-2015-073138], funded by the Spanish
State Research Agency.


\begin{thebibliography}{99}


\bibitem{Gisbert:2017vvj}
  H.~Gisbert and A.~Pich,
  ``Direct CP violation in $K^0\to\pi\pi$: Standard Model Status'',
  Rept.\ Prog.\ Phys.\  {\bf 81} (2018) no.7,  076201
  [arXiv:1712.06147 [hep-ph]].

 
\bibitem{Batley:2002gn}
  J.~R.~Batley {\it et al.}  [NA48 Collaboration],
  ``A Precision measurement of direct CP violation in the decay of neutral kaons into two pions'',
  {\it Phys. Lett.} B {\bf 544} (2002) 97
  [hep-ex/0208009].
  
\bibitem{Lai:2001ki}
  A.~Lai {\it et al.}  [NA48 Collaboration],
  ``A Precise measurement of the direct CP violation parameter $\mathrm{Re}(\varepsilon'/\varepsilon)$'',
  {\it Eur. Phys. J.} C {\bf 22} (2001) 231
  [hep-ex/0110019].
  
\bibitem{Fanti:1999nm}
  V.~Fanti {\it et al.}  [NA48 Collaboration],
  ``A New measurement of direct CP violation in two pion decays of the neutral kaon'',
  Phys.\ Lett.\ B
  {\bf 465} (1999) 335
  [hep-ex/9909022].
  
\bibitem{Barr:1993rx}
  G.~D.~Barr {\it et al.}  [NA31 Collaboration],
  ``A New measurement of direct CP violation in the neutral kaon system'',
  {\it Phys. Lett.} B {\bf 317} (1993) 233.
  
\bibitem{Burkhardt:1988yh}
  H.~Burkhardt {\it et al.}  [NA31 Collaboration],
  ``First Evidence for Direct CP Violation'',
  Phys.\ Lett.\ B
  {\bf 206} (1988) 169.
  
\bibitem{Abouzaid:2010ny}
  E.~Abouzaid {\it et al.}  [KTeV Collaboration],
  ``Precise Measurements of Direct CP Violation, CPT Symmetry, and Other Parameters in the Neutral Kaon System'',
  {\it Phys. Rev.} D {\bf 83} (2011) 092001
  [arXiv:1011.0127 [hep-ex]].
  
\bibitem{AlaviHarati:2002ye}
  A.~Alavi-Harati {\it et al.}  [KTeV Collaboration],
  ``Measurements of direct CP violation, CPT symmetry, and other parameters in the neutral kaon system'',
  Phys.\ Rev.\ D
{\bf 67} (2003) 012005
   [{\it Erratum-ibid.} D {\bf 70} (2004) 079904]
  [hep-ex/0208007].
  
\bibitem{AlaviHarati:1999xp}
  A.~Alavi-Harati {\it et al.}  [KTeV Collaboration],
  ``Observation of direct CP violation in $K_{S,L}\to\pi\pi$ decays'',
  {\it Phys. Rev. Lett.}  {\bf 83} (1999) 22
  [hep-ex/9905060].
  
\bibitem{Gibbons:1993zq} L.~K.~Gibbons {\it et al.} [E731 Collaboration],
  ``Measurement of the CP violation parameter $\mathrm{Re} (\varepsilon'/\varepsilon)$'',
  {\it Phys. Rev. Lett.}  {\bf 70} (1993) 1203.
  
\bibitem{Buras:1993dy}
  A.~J.~Buras, M.~Jamin and M.~E.~Lautenbacher,
  ``The Anatomy of $\varepsilon' / \varepsilon$ beyond leading logarithms with improved hadronic matrix elements'',
  Nucl.\ Phys.\ B {\bf 408} (1993) 209
  [hep-ph/9303284].
  
\bibitem{Buras:1996dq}
  A.~J.~Buras, M.~Jamin and M.~E.~Lautenbacher,
  ``A 1996 analysis of the CP violating ratio $\varepsilon' / \varepsilon$'',
  {\it Phys. Lett.} B {\bf 389} (1996) 749
  [hep-ph/9608365].
  
\bibitem{Bosch:1999wr} S.~Bosch {\it et al.},
  ``Standard model confronting new results for $\varepsilon' / \varepsilon$'',
  Nucl.\ Phys.\ B {\bf 565} (2000) 3
  [hep-ph/9904408].
  
\bibitem{Buras:2000qz} A.~J.~Buras {\it et al.},
  ``$\varepsilon' / \varepsilon$ and rare K and B decays in the MSSM'',
  Nucl.\ Phys.\ B {\bf 592} (2001) 55
  [hep-ph/0007313].
  
\bibitem{Ciuchini:1995cd}   M.~Ciuchini {\it et al.},
  ``An Upgraded analysis of $\varepsilon' / \varepsilon$ at the next-to-leading order'',
  {\it Z. Phys.} C {\bf 68} (1995) 239
  [hep-ph/9501265].
  
\bibitem{Ciuchini:1992tj}  M.~Ciuchini {\it et al.},
  ``$\varepsilon' / \varepsilon$ at the Next-to-leading order in QCD and QED'',
  {\it Phys. Lett.} B {\bf 301} (1993) 263
  [hep-ph/9212203].
 
\bibitem{Pallante:1999qf}
  E.~Pallante and A.~Pich,
  ``Strong enhancement of $\varepsilon' / \varepsilon$ through final state interactions'',
  {\it Phys. Rev. Lett.}  {\bf 84} (2000) 2568
  [hep-ph/9911233].

\bibitem{Pallante:2000hk}
  E.~Pallante and A.~Pich,
  ``Final state interactions in kaon decays'',
  {\it Nucl. Phys.} B {\bf 592} (2001) 294
  [hep-ph/0007208].
    
\bibitem{Pallante:2001he}
  E.~Pallante, A.~Pich and I.~Scimemi,
  ``The Standard model prediction for $\varepsilon' / \varepsilon$'',
  {\it Nucl. Phys.} B {\bf 617} (2001) 441
 [hep-ph/0105011].

  
\bibitem{Blum:2015ywa}
  T.~Blum {\it et al.},
  ``$K \rightarrow \pi\pi$ $\Delta I=3/2$ decay amplitude in the continuum limit'',
  Phys.\ Rev.\ D {\bf 91} (2015) no.7,  074502
  [arXiv:1502.00263 [hep-lat]].
  
\bibitem{Blum:2011ng} 
  T.~Blum {\it et al.}  [RBC and UKQCD Collaborations],
  ``The $K\to(\pi\pi)_{I=2}$ Decay Amplitude from Lattice QCD'',
  Phys.\ Rev.\ Lett.\
  {\bf 108} (2012) 141601;
  [arXiv:1111.1699 [hep-lat]];
  
\bibitem{Blum:2012uk}
T.~Blum {\it et al.} [RBC and UKQCD Collaborations],
  ``Lattice determination of the $K \to (\pi\pi)_{I=2}$ Decay Amplitude $A_2$'',
  {\it Phys. Rev.} D {\bf 86} (2012) 074513
  [arXiv:1206.5142 [hep-lat]].
  
\bibitem{Boyle:2012ys}
  P.~A.~Boyle {\it et al.}  [RBC and UKQCD Collaborations],
  ``Emerging understanding of the $\Delta I = 1/2$ Rule from Lattice QCD'',
  {\it Phys. Rev. Lett.}  {\bf 110} (2013) 152001
  [arXiv:1212.1474 [hep-lat]].
  
\bibitem{Pich:1995qp}
  A.~Pich and E.~de Rafael,
  ``Weak K amplitudes in the chiral and $1/N_c$ expansions'',
  {\it Phys. Lett.} B {\bf 374} (1996) 186
  [hep-ph/9511465].
  
\bibitem{Pich:1990mw}
  A.~Pich and E.~de Rafael,
  ``Four quark operators and nonleptonic weak transitions'',
  {\it Nucl. Phys.} B {\bf 358} (1991) 311.
  
\bibitem{Jamin:1994sv}
  M.~Jamin and A.~Pich,
  ``QCD corrections to inclusive $\Delta S = 1,2$ transitions at the next-to-leading order'',
  {\it Nucl. Phys.} B {\bf 425} (1994) 15
  [hep-ph/9402363].
  
\bibitem{Bertolini:1997ir} S.~Bertolini {\it et al.},
  ``The $\Delta I = 1/2$ rule and $B_K$ at $O (p^4)$ in the chiral expansion'',
  {\it Nucl. Phys.} B {\bf 514} (1998) 63
  [hep-ph/9705244].
  
\bibitem{Antonelli:1995nv} V.~Antonelli {\it et al.},
  ``The $\Delta S = 1$ weak chiral lagrangian as the effective theory of the chiral quark model'',
  Nucl.\ Phys.\ B {\bf 469} (1996) 143
   [hep-ph/9511255];
  
\bibitem{Antonelli:1995gw} V.~Antonelli {\it et al.},
  ``The $\Delta I = 1/2$ selection rule'',
  Nucl.\ Phys.\ B {\bf 469} (1996) 181
  [hep-ph/9511341].
  
\bibitem{Hambye:1998sma} T.~Hambye {\it et al.},
  ``$1 / N_c$ corrections to the hadronic matrix elements of $Q_6$ and $Q_8$ in $K \to \pi \pi$ decays'',
  Phys.\ Rev.\ D {\bf 58} (1998) 014017
  [hep-ph/9802300].
  
\bibitem{Bardeen:1986vz}
  W.~A.~Bardeen, A.~J.~Buras and J.~M.~Gerard,
  ``A Consistent Analysis of the $\Delta I = 1/2$ Rule for K Decays'',
  {\it Phys. Lett.} B {\bf 192} (1987) 138.
  
\bibitem{Buras:2014maa}
  A.~J.~Buras, J.~M.~Gerard and W.~A.~Bardeen,
  ``Large $N$ Approach to Kaon Decays and Mixing 28 Years Later: $\Delta I = 1/2$ Rule, $\hat B_K$ and $\Delta M_K$'',
  {\it Eur. Phys. J.} C {\bf 74} (2014) 2871
  [arXiv:1401.1385 [hep-ph]].
  
\bibitem{Bijnens:1998ee}
  J.~Bijnens and J.~Prades,
  ``The Delta I = 1/2 rule in the chiral limit'',
  {\it JHEP} {\bf 9901} (1999) 023
  [hep-ph/9811472].
  
  
\bibitem{Bai:2015nea}
  Z.~Bai {\it et al.} [RBC and UKQCD Collaborations],
  ``Standard Model Prediction for Direct CP Violation in $K\to\pi\pi$ Decay'',
  Phys.\ Rev.\ Lett.\  {\bf 115} (2015) no.21,  212001
  [arXiv:1505.07863 [hep-lat]].
  
\bibitem{Buras:2015xba}
  A.~J.~Buras and J.~M.~Gérard,
  ``Upper bounds on $\varepsilon'/\varepsilon$  parameters B$_{6}^{(1/2)}$ and B$_{8}^{(3/2)}$ from large N QCD and other news'',
  JHEP {\bf 1512} (2015) 008
  [arXiv:1507.06326 [hep-ph]].
  
\bibitem{Buras:2016fys}
  A.~J.~Buras and J.~M.~Gerard,
  ``Final state interactions in $K\rightarrow \pi \pi $ decays: $\Delta I=1/2$ rule vs. $\varepsilon '/\varepsilon $'',
  Eur.\ Phys.\ J.\ C {\bf 77} (2017) no.1,  10
  [arXiv:1603.05686 [hep-ph]].
  
\bibitem{Buras:2015yba} A.~J.~Buras {\it et al.},
 ``Improved anatomy of $\varepsilon '/\varepsilon $ in the Standard Model'',
  JHEP {\bf 1511} (2015) 202
  [arXiv:1507.06345 [hep-ph]].
  
\bibitem{Buras:2014sba}
  A.~J.~Buras, F.~De Fazio and J.~Girrbach,
  ``$\Delta I=1/2$ rule, $\varepsilon'/\varepsilon $ and $K\rightarrow \pi \nu \bar{\nu }$ in $Z' (Z)$ and $G' $ models with FCNC quark couplings'',
  Eur.\ Phys.\ J.\ C {\bf 74} (2014) no.7,  2950
  [arXiv:1404.3824 [hep-ph]].

\bibitem{Buras:2015yca}
  A.~J.~Buras, D.~Buttazzo and R.~Knegjens,
  ``$ K\to \pi \nu \overline{\nu} $ and $\varepsilon'/\varepsilon $ in simplified new physics models'',
  JHEP {\bf 1511} (2015) 166
  [arXiv:1507.08672 [hep-ph]].
  
\bibitem{Blanke:2015wba}
  M.~Blanke, A.~J.~Buras and S.~Recksiegel,
  ``Quark flavour observables in the Littlest Higgs model with T-parity after LHC Run 1'',
  Eur.\ Phys.\ J.\ C {\bf 76} (2016) no.4,  182
  [arXiv:1507.06316 [hep-ph]].

\bibitem{Buras:2015kwd}
  A.~J.~Buras and F.~De Fazio,
  ``$\varepsilon'/\varepsilon$ in 331 Models'',
  JHEP {\bf 1603} (2016) 010
  [arXiv:1512.02869 [hep-ph]].

\bibitem{Buras:2016dxz}
  A.~J.~Buras and F.~De Fazio,
  ``331 Models Facing the Tensions in $\Delta F=2$ Processes with the Impact on $\varepsilon^\prime/\varepsilon$, $B_s\to\mu^+\mu^-$ and $B\to K^*\mu^+\mu^-$'',
  JHEP {\bf 1608} (2016) 115
  [arXiv:1604.02344 [hep-ph]].

\bibitem{Buras:2015jaq}
  A.~J.~Buras,
  ``New physics patterns in $\varepsilon^\prime/\varepsilon$ and $\varepsilon_K$ with implications for rare kaon decays and $\Delta M_K$'',
  JHEP {\bf 1604} (2016) 071
  [arXiv:1601.00005 [hep-ph]].
      
\bibitem{Kitahara:2016otd}
  T.~Kitahara, U.~Nierste and P.~Tremper,
  ``Supersymmetric Explanation of CP Violation in $K\to \pi\pi$ Decays'',
  Phys.\ Rev.\ Lett.\  {\bf 117} (2016) no.9,  091802
  [arXiv:1604.07400 [hep-ph]].

\bibitem{Kitahara:2016nld}
  T.~Kitahara, U.~Nierste and P.~Tremper,
  ``Singularity-free next-to-leading order $\Delta$S = 1 renormalization group evolution and $\epsilon_K'/\epsilon_K$ in the Standard Model and beyond'',
  JHEP {\bf 1612} (2016) 078
  [arXiv:1607.06727 [hep-ph]].

\bibitem{Endo:2016aws} M.~Endo {\it et al.},
  ``Chargino contributions in light of recent $\epsilon'/\epsilon$'',
  Phys.\ Lett.\ B {\bf 762} (2016) 493
  [arXiv:1608.01444 [hep-ph]].

\bibitem{Endo:2016tnu} M.~Endo {\it et al.},
  ``Revisiting Kaon Physics in General $Z$ Scenario'',
  Phys.\ Lett.\ B {\bf 771} (2017) 37
  [arXiv:1612.08839 [hep-ph]].

\bibitem{Cirigliano:2016yhc} V.~Cirigliano {\it et al.},
  ``An $\epsilon'$ improvement from right-handed currents'',
  Phys.\ Lett.\ B {\bf 767} (2017) 1
  [arXiv:1612.03914 [hep-ph]].

\bibitem{Alioli:2017ces} S.~Alioli {\it et al.},
  ``Right-handed charged currents in the era of the Large Hadron Collider'',
  JHEP {\bf 1705} (2017) 086
  [arXiv:1703.04751 [hep-ph]].
  
\bibitem{Bobeth:2016llm} C.~Bobeth {\it et al.},
  ``Patterns of Flavour Violation in Models with Vector-Like Quarks'',
  JHEP {\bf 1704} (2017) 079
  [arXiv:1609.04783 [hep-ph]].
    
\bibitem{Bobeth:2017xry} C.~Bobeth {\it et al.},
  ``Yukawa enhancement of $Z$-mediated new physics in $\Delta S = 2$ and $\Delta B = 2$ processes'',
  JHEP {\bf 1707} (2017) 124
  [arXiv:1703.04753 [hep-ph]].

\bibitem{Crivellin:2017gks} A.~Crivellin {\it et al.},
  ``$K\to \pi \nu\overline{\nu}$ in the MSSM in light of the $\epsilon^{\prime}_K/\epsilon_K$ anomaly'',
  Phys.\ Rev.\ D {\bf 96} (2017) no.1,  015023
  [arXiv:1703.05786 [hep-ph]].


\bibitem{Chobanova:2017rkj} V.~Chobanova {\it et al.},
  ``Probing SUSY effects in $K_S^0\rightarrow\mu^+\mu^-$'',
  JHEP {\bf 1805} (2018) 024
  [arXiv:1711.11030 [hep-ph]].

\bibitem{Bobeth:2017ecx}
  C.~Bobeth and A.~J.~Buras,
  ``Leptoquarks meet $\varepsilon'/\varepsilon$ and rare Kaon processes'',
  JHEP {\bf 1802} (2018) 101
  [arXiv:1712.01295 [hep-ph]].

\bibitem{Endo:2017ums} M.~Endo {\it et al.},
  ``Gluino-mediated electroweak penguin with flavor-violating trilinear couplings'',
  JHEP {\bf 1804} (2018) 019
  [arXiv:1712.04959 [hep-ph]].

\bibitem{Chen:2018dfc}
  C.~H.~Chen and T.~Nomura,
  ``$\epsilon_K$ and $\epsilon'/\epsilon$ in a diquark model'',
  arXiv:1808.04097 [hep-ph].
  
\bibitem{Aebischer:2018quc} J.~Aebischer {\it et al.},
  ``Master formula for $\varepsilon'/\varepsilon$ beyond the Standard Model'',
  arXiv:1807.02520 [hep-ph].
  
\bibitem{Aebischer:2018rrz}
  J.~Aebischer, A.~J.~Buras and J.~M.~Gérard,
  ``BSM Hadronic Matrix Elements for $\epsilon'/\epsilon$ and $K\to\pi\pi$ Decays in the Dual QCD Approach'',
  arXiv:1807.01709 [hep-ph].
  
\bibitem{Haba:2018rzf}
  N.~Haba, H.~Umeeda and T.~Yamada,
  ``Direct CP Violation in Cabibbo-Favored Charmed Meson Decays and $\epsilon'/\epsilon$ in $SU(2)_L\times SU(2)_R\times U(1)_{B-L}$ Model'',
  arXiv:1806.03424 [hep-ph].
  
\bibitem{Matsuzaki:2018jui}
  S.~Matsuzaki, K.~Nishiwaki and K.~Yamamoto,
  ``Simultaneous interpretation of $K$ and $B$ anomalies in terms of chiral-flavorful vectors'',
  arXiv:1806.02312 [hep-ph].
  
\bibitem{Chen:2018vog}
  C.~H.~Chen and T.~Nomura,
  ``$\epsilon'/\epsilon$ from charged-Higgs-induced gluonic dipole operators'',
  arXiv:1805.07522 [hep-ph].
  
\bibitem{Chen:2018ytc}
  C.~H.~Chen and T.~Nomura,
  ``Re($\varepsilon_{K}^{'} /\varepsilon_{K}$) and $ K\to \pi \nu \overline{\nu} $ in a two-Higgs doublet model'',
  JHEP {\bf 1808} (2018) 145
  [arXiv:1804.06017 [hep-ph]].
  
\bibitem{Haba:2018byj}
  N.~Haba, H.~Umeeda and T.~Yamada,
  ``$\epsilon'/\epsilon$ Anomaly and Neutron EDM in $SU(2)_L\times SU(2)_R\times U(1)_{B-L}$ model with Charge Symmetry'',
  JHEP {\bf 1805} (2018) 052
  [arXiv:1802.09903 [hep-ph]].


\bibitem{Kelly:CKM2018}
C. Kelly, ``Progress in lattice in the kaon system'', talk at CKM 2018 (Heidelberg, September 17th).

  
\bibitem{Cirigliano:2003nn}
  V.~Cirigliano, A.~Pich, G.~Ecker and H.~Neufeld,
  ``Isospin violation in $\varepsilon'$'',
  {\it Phys. Rev. Lett.}  {\bf 91} (2003) 162001
  [hep-ph/0307030].
  
\bibitem{Cirigliano:2003gt}
  V.~Cirigliano, G.~Ecker, H.~Neufeld and A.~Pich,
  ``Isospin breaking in $K\to\pi\pi$ decays'',
  {\it Eur. Phys. J.} C {\bf 33} (2004) 369
  [hep-ph/0310351];
  
\bibitem{Cirigliano:2009rr}
  V.~Cirigliano, G.~Ecker and A.~Pich,
  ``Reanalysis of pion pion phase shifts from $K\to\pi\pi$ decays'',
  {\it Phys. Lett.} B {\bf 679} (2009) 445
  [arXiv:0907.1451 [hep-ph]].

\bibitem{Aoki:2016frl}
  S.~Aoki {\it et al.},
  ``Review of lattice results concerning low-energy particle physics'',
  Eur.\ Phys.\ J.\ C {\bf 77} (2017) no.2,  112
  [arXiv:1607.00299 [hep-lat]].

\bibitem{Ecker:1988te}
  G.~Ecker, J.~Gasser, A.~Pich and E.~de Rafael,
  ``The Role of Resonances in Chiral Perturbation Theory'',
  Nucl.\ Phys.\ B {\bf 321} (1989) 311.

\bibitem{Ecker:1989yg}
  G.~Ecker, J.~Gasser, H.~Leutwyler, A.~Pich and E.~de Rafael,
  ``Chiral Lagrangians for Massive Spin 1 Fields'',
  Phys.\ Lett.\ B {\bf 223} (1989) 425.

\bibitem{Pich:2002xy}
  A.~Pich,
  ``Colorless mesons in a polychromatic world'',
Proc. Int. Workshop on {\it Phenomenology of Large $N_C$ QCD} (Tempe, Arizona, 2002), ed. R. F. Lebed, Proc. Institute for Nuclear Theory -- Vol. 12 (World Scientific, Singapore, 2002), p.~239  [hep-ph/0205030].

\bibitem{Cirigliano:2006hb}
  V.~Cirigliano, G.~Ecker, M.~Eidemuller, R.~Kaiser, A.~Pich and J.~Portol\'es,
  ``Towards a consistent estimate of the chiral low-energy constants'',
  Nucl.\ Phys.\ B {\bf 753} (2006) 139
  [hep-ph/0603205].

\bibitem{Kaiser:2007zz}
  R.~Kaiser,
  Nucl.\ Phys.\ Proc.\ Suppl.\  {\bf 174} (2007) 97.
  
\bibitem{Cirigliano:2004ue}
  V.~Cirigliano, G.~Ecker, M.~Eidemuller, A.~Pich and J.~Portol\'es,
  ``$\langle VAP\rangle$  Green function in the resonance region'',
  Phys.\ Lett.\ B {\bf 596} (2004) 96
  [hep-ph/0404004].

\bibitem{Cirigliano:2005xn}
  V.~Cirigliano, G.~Ecker, M.~Eidemuller, R.~Kaiser, A.~Pich and J.~Portol\'es,
  ``The  Green function and SU(3) breaking in $K_{l3}$ decays'',
  JHEP {\bf 0504} (2005) 006
  [hep-ph/0503108].

\bibitem{RuizFemenia:2003hm}
  P.~D.~Ruiz-Femenia, A.~Pich and J.~Portol\'es,
  ``Odd intrinsic parity processes within the resonance effective theory of QCD'',
  JHEP {\bf 0307} (2003) 003
  [hep-ph/0306157].

\bibitem{Jamin:2004re}
  M.~Jamin, J.~A.~Oller and A.~Pich,
  ``Order $p^{6}$ chiral couplings from the scalar $K \pi$ form-factor'',
  JHEP {\bf 0402} (2004) 047
  [hep-ph/0401080].
  
\bibitem{Rosell:2004mn}
  I.~Rosell, J.~J.~Sanz-Cillero and A.~Pich,
  ``Quantum loops in the resonance chiral theory: The Vector form-factor'',
  JHEP {\bf 0408} (2004) 042
  [hep-ph/0407240].

\bibitem{Rosell:2006dt}
  I.~Rosell, J.~J.~Sanz-Cillero and A.~Pich,
  ``Towards a determination of the chiral couplings at NLO in $1/N_C$: $L^r_8(\mu)$'',
  JHEP {\bf 0701} (2007) 039
  [hep-ph/0610290].

\bibitem{Pich:2008jm}
  A.~Pich, I.~Rosell and J.~J.~Sanz-Cillero,
  ``Form-factors and current correlators: Chiral couplings $L_{10}^r(\mu)$ and $C_{87}^r(\mu)$ at NLO in $1/N_C$'',
  JHEP {\bf 0807} (2008) 014
  [arXiv:0803.1567 [hep-ph]].
 
\bibitem{GonzalezAlonso:2008rf}
  M.~Gonzalez-Alonso, A.~Pich and J.~Prades,
  ``Determination of the Chiral Couplings $L_{10}$ and $C_{87}$ from Semileptonic Tau Decays'',
  Phys.\ Rev.\ D {\bf 78} (2008) 116012
  [arXiv:0810.0760 [hep-ph]].
  
\bibitem{Pich:2010sm}
  A.~Pich, I.~Rosell and J.~J.~Sanz-Cillero,
  ``The vector form factor at the next-to-leading order in $1/N_C$: chiral couplings $L_9(\mu)$ and $C_{88}(\mu) - C_{90}(\mu)$'',
  JHEP {\bf 1102} (2011) 109
  [arXiv:1011.5771 [hep-ph]].

\bibitem{Bijnens:2014lea}
  J.~Bijnens and G.~Ecker,
  ``Mesonic low-energy constants'',
  Ann.\ Rev.\ Nucl.\ Part.\ Sci.\  {\bf 64} (2014) 149
  [arXiv:1405.6488 [hep-ph]].

\bibitem{Rodriguez-Sanchez:2016jvw}
  M.~González-Alonso, A.~Pich and A.~Rodríguez-Sánchez,
  ``Updated determination of chiral couplings and vacuum condensates from hadronic $\tau$ decay data'',
  Phys.\ Rev.\ D {\bf 94} (2016) no.1,  014017
  [arXiv:1602.06112 [hep-ph]].


\bibitem{Ananthanarayan:2017qmx} B.~Ananthanarayan {\it et al.},
  Phys.\ Rev.\ D {\bf 97} (2018) no.9,  091502
  [arXiv:1711.11328 [hep-ph]].

  
\bibitem{Antonelli:2010yf} M.~Antonelli {\it et al.},
  {\it Eur. Phys. J.} C {\bf 69} (2010) 399.
  [arXiv:1005.2323 [hep-ph]].
  
\bibitem{Buchalla:1995vs}
  G.~Buchalla, A.~J.~Buras and M.~E.~Lautenbacher,
  ``Weak decays beyond leading logarithms'',
  {\it Rev. Mod. Phys.}  {\bf 68} (1996) 1125
  [hep-ph/9512380].
  
\bibitem{Buras:1991jm} A.~J.~Buras {\it et al.},
  ``Effective Hamiltonians for $\Delta S = 1$ and $\Delta B = 1$ nonleptonic decays beyond the leading logarithmic approximation'',
  Nucl.\ Phys.\ B {\bf 370} (1992) 69
   [Addendum: Nucl.\ Phys.\ B {\bf 375} (1992) 501].
  
\bibitem{Buras:1992tc} A.~J.~Buras {\it et al.},
  {\it Nucl. Phys.} B {\bf 400} (1993) 37
  [hep-ph/9211304].

\bibitem{Buras:1992zv}
  A.~J.~Buras, M.~Jamin and M.~E.~Lautenbacher,
  Nucl.\ Phys.\ B {\bf 400} (1993) 75
  [hep-ph/9211321].

\bibitem{Ciuchini:1993vr} M.~Ciuchini {\it et al.},
  ``The $\Delta S = 1$ effective Hamiltonian including next-to-leading order QCD and QED corrections'',
  {\it Nucl. Phys.} B {\bf 415} (1994) 403
  [hep-ph/9304257].
  
\bibitem{Buras:1999st}
  A.~J.~Buras, P.~Gambino and U.~A.~Haisch,
  Nucl.\ Phys.\ B {\bf 570} (2000) 117
  [hep-ph/9911250].

\bibitem{Gorbahn:2004my}
  M.~Gorbahn and U.~Haisch,
  Nucl.\ Phys.\ B {\bf 713} (2005) 291
  [hep-ph/0411071].

\bibitem{Cerda-Sevilla:2016yzo} M.~Cerdà-Sevilla {\it et al.},
  J.\ Phys.\ Conf.\ Ser.\  {\bf 800} (2017) no.1,  012008
  [arXiv:1611.08276 [hep-ph]].
  
\bibitem{Weinberg:1978kz}
  S.~Weinberg,
  ``Phenomenological Lagrangians'',
  Physica A {\bf 96} (1979) 327.

\bibitem{Gasser:1983yg}
  J.~Gasser and H.~Leutwyler,
  ``Chiral Perturbation Theory to One Loop'',
  Annals Phys.\  {\bf 158} (1984) 142.

\bibitem{Gasser:1984gg}
  J.~Gasser and H.~Leutwyler,
  ``Chiral Perturbation Theory: Expansions in the Mass of the Strange Quark'',
  Nucl.\ Phys.\ B {\bf 250} (1985) 465.


\bibitem{Cirigliano:2011ny}
  V.~Cirigliano, G.~Ecker, H.~Neufeld, A.~Pich and J.~Portoles,
  ``Kaon Decays in the Standard Model'',
  Rev.\ Mod.\ Phys.\  {\bf 84} (2012) 399
  [arXiv:1107.6001 [hep-ph]].


  
\bibitem{Buras:1987wc}
  A.~J.~Buras and J.~M.~Gerard,
  Phys.\ Lett.\ B {\bf 192} (1987) 156.
  
\bibitem{Buras:1985yx}
  A.~J.~Buras and J.~M.~Gerard,
  ``$1/N$ Expansion for Kaons'',
  Nucl.\ Phys.\ B {\bf 264} (1986) 371.
  
\bibitem{Bardeen:1986uz}
  W.~A.~Bardeen, A.~J.~Buras and J.~M.~Gerard,
  ``The $K \to\pi \pi$ Decays in the Large n Limit: Quark Evolution'',
  Nucl.\ Phys.\ B {\bf 293} (1987) 787.
  

\bibitem{Bertolini:1995tp}
  S.~Bertolini, J.~O.~Eeg and M.~Fabbrichesi,
  ``A New estimate of $\varepsilon' / \varepsilon$'',
  Nucl.\ Phys.\ B {\bf 476} (1996) 225
  [hep-ph/9512356].
  
\bibitem{Bertolini:1997nf}
  S.~Bertolini, J.~O.~Eeg, M.~Fabbrichesi and E.~I.~Lashin,
  ``$\varepsilon' / \varepsilon$ at $O(p^4)$ in the chiral expansion'',
  Nucl.\ Phys.\ B {\bf 514} (1998) 93
  [hep-ph/9706260].
  
\bibitem{Bertolini:1998vd}
  S.~Bertolini, M.~Fabbrichesi and J.~O.~Eeg,
  ``Theory of the CP violating parameter $\varepsilon' / \varepsilon$'',
  Rev.\ Mod.\ Phys.\  {\bf 72} (2000) 65
  [hep-ph/9802405].

\bibitem{Bertolini:2000dy}
  S.~Bertolini, J.~O.~Eeg and M.~Fabbrichesi,
  ``An Updated analysis of $\varepsilon' / \varepsilon$ in the standard model with hadronic matrix elements from the chiral quark model'',
  Phys.\ Rev.\ D {\bf 63} (2001) 056009
  [hep-ph/0002234].

\bibitem{Hambye:1999yy}
  T.~Hambye, G.~O.~Kohler, E.~A.~Paschos and P.~H.~Soldan,
  ``Analysis of $\varepsilon' / \varepsilon$ in the $1 / N_C$ expansion'',
  Nucl.\ Phys.\ B {\bf 564} (2000) 391
  [hep-ph/9906434].





\bibitem{VincenzoEtAl} V.~Cirigliano {\it et. al.}, work in progress.

\bibitem{Antonio} A.~Rodríguez-Sánchez and A.~Pich, work in progress.

  
\bibitem{Buchler:2001nm}
  M.~Buchler, G.~Colangelo, J.~Kambor and F.~Orellana,
  ``Dispersion relations and soft pion theorems for $K \to\pi \pi$'',
  Phys.\ Lett.\ B {\bf 521} (2001) 22
  [hep-ph/0102287].
  
\bibitem{Buchler:2005xn}
  M.~Buchler,
  ``The Chiral logs of the $K \to\pi \pi$ amplitude'',
  Phys.\ Lett.\ B {\bf 633} (2006) 497
  [hep-ph/0511087].
    
  
\bibitem{Bijnens:2000im}
  J.~Bijnens and J.~Prades,
  ``$\varepsilon_K' / \varepsilon_K$ in the chiral limit'',
  JHEP {\bf 0006} (2000) 035
  [hep-ph/0005189].
  
\bibitem{Hambye:2003cy}
  T.~Hambye, S.~Peris and E.~de Rafael,
  ``$\Delta I = 1/2$ and $\varepsilon' / \varepsilon$ in large $N_C$ QCD'',
  JHEP {\bf 0305} (2003) 027
  [hep-ph/0305104].

\bibitem{Feng:2017voh}
  X.~Feng,
  ``Recent progress in applying lattice QCD to kaon physics'',
  EPJ Web Conf.\  {\bf 175} (2018) 01005
  [arXiv:1711.05648 [hep-lat]].


\end{thebibliography}
\end{document}